\documentclass[11pt,superscriptaddress,showpacs,aps,preprint]{revtex4}

\usepackage{amsmath}
\usepackage{amssymb}
\usepackage{graphicx}
\usepackage{color}
\usepackage{subfigure}

\makeatletter

\usepackage{slashed}

\newcommand{\be}{\begin{equation}}
	\newcommand{\ee}{\end{equation}}
\newcommand{\en}{\end{equation}}
\newcommand{\ba}{\begin{eqnarray}}
\newcommand{\ea}{\end{eqnarray}}
\newcommand{\bea}{\begin{eqnarray}}
\newcommand{\eea}{\end{eqnarray}}

\newcommand{\bq}{\begin{eqnarray}}
\newcommand{\eq}{\end{eqnarray}}

\makeatother

\begin{document}

\title {One-loop calculations in CPT-even Lorentz-breaking scalar QED}

\author{A. P. Ba\^eta Scarpelli}
\email{scarpelli@cefetmg.br}
\affiliation{Centro Federal de Educa\c{c}\~ao Tecnol\'ogica - MG \\
	Avenida Amazonas, 7675 - 30510-000 - Nova Gameleira - Belo Horizonte
	-MG - Brazil}

\author{J. C. C. Felipe}
\email{jean.cfelipe@ufvjm.edu.br}
\affiliation{Instituto de Engenharia, Ci\^{e}ncia e Tecnologia, Universidade Federal dos Vales do Jequitinhonha e Mucuri, Avenida Um, 4050 - 39447-790 - Cidade Universit\'{a}ria - Jana\'{u}ba - MG- Brazil}

\author{L. C. T. Brito}
\email{lcbrito@dfi.ufla.br}
\affiliation{Departamento de F\'{i}sica, Universidade Federal de Lavras, Caixa Postal 3037,
	37200-000, Lavras, MG, Brasil}

\author{A. Yu. Petrov}
\email{petrov@fisica.ufpb.br}
\affiliation{Departamento de F\'{i}sica, Universidade Federal da Paraíba, Caixa Postal 5008,
	58051-970 Jo\~{a}o Pessoa, Para\'{i}ba, Brazil}


\begin{abstract}
	In this paper, we study a CPT-even Lorentz-breaking extension of the scalar QED. For this theory, we calculate the one-loop lower-order contributions in the Lorentz-violating parameters to the two-point functions of scalar and gauge fields. We found that the two background tensors, coming from the two sectors (scalar and gauge) are mixed in the one-loop corrections both in finite and divergent parts. This shows that these two Lorentz-breaking terms cannot be studied in an isolated form. Besides, the results in the gauge sector are confirmed to be transversal.
	
\end{abstract}

\maketitle

\section{Introduction}
\label{I}

The formulation of the Lorentz-violating (LV) extension of Standard Model (SME), presented in \cite{kostelecky1,kostelecky2}, brought interest to perturbative calculations in several Lorentz symmetry breaking versions of well-known field theory models. One of the motivations for such calculations consists in the development of a scheme from which the desired LV terms arise as quantum corrections to some fundamental theory where vector (gauge) and scalar fields are coupled to some spinor field. This scheme has been followed in a number of papers beginning from \cite{JK}, in which it has been applied to generate the Carroll-Field-Jackiw (CFJ) term as an one-loop correction, and, in further papers, finite aether and higher-derivative terms were generated in the gauge sector of the theory (for a review on this methodology and discussion of many examples, see \cite{ourrev}).

On the experimental side, the search for possible effects of Lorentz violation is still a crucial task since they may present imprints of quantum gravity at low energy \cite{Kostelecky:1988zi,Doplicher:1994tu,Bojowald:2004bb,Amelino-Camelia:2008aez}. Experiments based on multimessenger astronomy, for instance, bring us new possibilities to test different models that incorporate Lorentz violation \cite{Addazi:2021xuf}. In particular, from the perspective of the Standard Model Extension (SME)  program, proposed tests involve gravitational waves \cite{Kostelecky:2016kfm,Kostelecky:2016kkn,Schreck:2016qiz}, cosmic rays \cite{Kostelecky:2015dpa,Altschul:2008qg}, and astrophysical tests with neutrinos and gamma-ray photons \cite{Bertolami:1999da}. Fortunately, there is an extensive compilation of constraints on the values of coefficients for Lorentz and CPT violation in the SME \cite{Kostelecky:2008ts}, permitting quantifying Lorentz violation from the different sectors of the SME. From the Data Tables \cite{Kostelecky:2008ts},  we can see that although the coefficients in the SME are very small, they cannot be disregarded. In particular, for the dimensionless coefficients in which we are interested in this paper, we have that $c^{\mu\nu}$ (scalar sector) is about $ 10^{-16}$ and $\kappa_{\mu\nu\lambda\rho}$  (photon sector) is about $10^{-17}$.

Another motivation is the study of the various LV theories on their own basis, such as the renormalization process, the calculation of effective potentials and the treatment of other issues of quantum field theory. For minimal LV extensions of spinorial QED, many results have been obtained in this context. For example, renormalization has been carried out, with the corresponding obtainment of the renormalization group equations in the first \cite{Kost2001a} and the second \cite{Scarp1,Scarp2} orders in LV parameters. 

At the same time, the LV extensions of the {\it scalar} QED are studied to a much lesser extent, mostly within the Higgs mechanism context. In this context, the most important studies were performed in \cite{AltHiggs}, where tree-level aspects of the Higgs mechanism  for the CPT-even LV QED were considered, and the prescription for the Faddeev-Popov quantization in this theory was formulated. In \cite{Scarp2013}, the Higgs mechanism was studied in the scalar QED with an additive CPT-odd CFJ term in the one-loop approximation. Also, in \cite{effpotLV} the effective potential for LV extensions of scalar QED was calculated. However, no other study of perturbative aspects of a CPT-even LV scalar QED was performed up to now. In this paper, we intend to begin filling this gap.


Explicitly, within the present study, we consider the CPT-even LV extension of the scalar QED originally introduced in \cite{Hel}, with Lorentz-violating terms added both in scalar and gauge sectors, and perform the one-loop calculations of divergent and finite corrections to the two-point functions of gauge and scalar fields in the first order in LV parameters. We expect that our results can be applied to studies of the Higgs mechanism in LV scalar QED.

The structure of the paper looks like follows:  in the Section II, we give the general description of our model and list relevant Feynman diagrams; the results of calculations are presented in Section III; in Section IV, we present our conclusions; we also present an appendix with a brief revision of Implicit Regularization, which is the procedure used in the calculations.

\section{Our model and lower corrections}

We start with the CPT-even QED with the Lagrangian
\bea
\label{lag1}
{\cal L}=(D_{\mu}\phi)^*(\eta^{\mu\nu}+c^{\mu\nu})(D_{\nu}\phi)-m^2\phi^*\phi-\frac{1}{4}F_{\mu\nu}F^{\mu\nu}+\frac{1}{4}\kappa_{\mu\nu\lambda\rho}F^{\mu\nu}F^{\lambda\rho}.
\eea
Here, the Lorentz symmetry breaking is introduced both in scalar and gauge sectors through additive terms proportional to the background tensors $c^{\mu\nu}$ and $\kappa^{\mu\nu\lambda\rho}$. Both CPT-even contributions have been introduced originally in \cite{kostelecky1,kostelecky2} and imply in a renormalizable theory since both $c^{\mu\nu}$ and $\kappa_{\mu\nu\lambda\rho}$ are dimensionless. As usual, we require $c^{\mu\nu}$ to be symmetric and traceless, and $\kappa_{\mu\nu\lambda\rho}$ to have the same symmetry as the Riemann curvature tensor. The tensor $c^{\mu\nu}$ of the present paper should not be confused with the one normally introduced in the fermionic sector of Standard Model Extension. Actually, what we call $c^{\mu\nu}$ is the $\kappa^{\mu\nu}_{\phi\phi}$ tensor of \cite{kostelecky2}, which must have a symmetric real part and an antisymmetric imaginary part. Here we consider a real $c^{\mu\nu}$, so that it must be symmetric.

\begin{figure}[h]
	\begin{center}
		\includegraphics[scale=0.9]{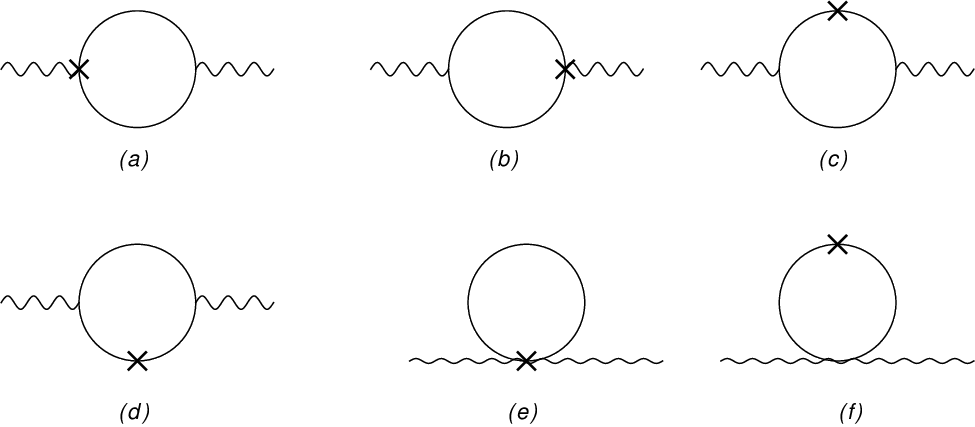}
		\caption{Diagrammatic representation of the one-loop two-point function of the gauge field at first-order in the Lorentz violation parameters. The wavy and solid lines represent the photon and scalar propagators, respectively, and the crosses indicate insertions of LV parameters.}
		\label{Fig1}
	\end{center}
\end{figure}

We would like to study the one-loop corrections to the two point functions both of the scalar and gauge sectors, which are interesting investigations that have never been performed. Starting from Lagrangian density (\ref{lag1}) and considering just the first order in $c^{\mu\nu}$ parameters, to study the two-point function of the gauge field, we must obtain contributions of the diagrams depicted in Figure \ref{Fig1}.

Extracting the appropriated Feynman rules for the model of (\ref{lag1}), we obtain that the contributions of the diagrams with external gauge legs depicted in Figure \ref{Fig1} look like as:

\bea
\label{lag3}
I_{(a)}&=&-\frac{c^{\alpha\mu}}{2}\int\frac{d^4k}{(2\pi)^4}\frac{(2k+p)_{\alpha}(2k+p)_{\nu}}{[k^2-m^2][(k+p)^2-m^2]}A_{\mu}(-p)A^{\nu}(p);\nonumber \\
I_{(b)}&=&-\frac{c^{\alpha\nu}}{2}\int\frac{d^4k}{(2\pi)^4}\frac{(2k+p)_{\mu}(2k+p)_{\alpha}}{[k^2-m^2][(k+p)^2-m^2]}A^{\mu}(-p)A_{\nu}(p);\nonumber \\
I_{(c)}&=&\frac{c^{\alpha\beta}}{2}\int\frac{d^4k}{(2\pi)^4}\frac{(2k+p)_{\mu}(2k+p)_{\nu}k_{\alpha}k_{\beta}}{[k^2-m^2]^2[(k+p)^2-m^2]}A^{\mu}(-p)A^{\nu}(p); \nonumber \\
I_{(d)}&=&\frac{c^{\alpha\beta}}{2}\int\frac{d^4k}{(2\pi)^4}\frac{(2k+p)_{\mu}(2k+p)_{\nu}(k+p)_{\alpha}(k+p)_{\beta}}{[k^2-m^2][(k+p)^2-m^2]^2} A^{\mu}(-p)A^{\nu}(p); \nonumber\\
I_{(e)}&=& -\eta^{\mu\nu}c^{\alpha\beta}\int\frac{d^{4}k}{(2\pi)^{4}}\frac{k_{\alpha}k_{\beta}}{(k^{2}-m^{2})^2}A_\mu(-p)A_\nu(p); \nonumber\\
I_{(f)}&=& c^{\mu\nu}\int\frac{d^{4}k}{(2\pi)^{4}}\frac{1}{(k^{2}-m^{2})^2}A_\mu(-p)A_\nu(p).
\eea

\begin{figure}[h]
	\begin{center}
		\includegraphics[scale=0.9]{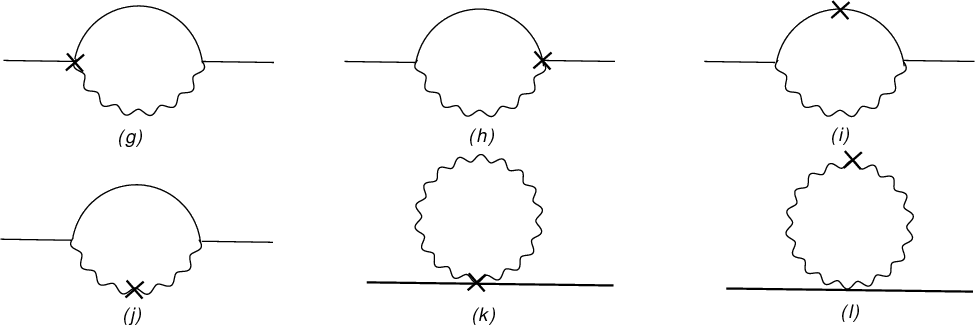}
		\caption{Diagrammatic representation of the scalar field two-point function at first-order in the Lorentz violation parameters. The wavy and solid lines represent the photon and scalar propagators, respectively, and the crosses indicate the insertions.}
		\label{Fig2}
	\end{center}
\end{figure}

Besides of the gauge sector, we must obtain the contributions with external matter legs. Here, we keep only first order in the LV parameters. So, for the terms in $c^{\mu\nu}$, the gauge propagator is not modified. For these contributions we have the diagrams depicted in Figure \ref{Fig2}. 

The first three diagrams yield the following amplitudes
\bea
I_{(g)}&=&-\eta^{\mu\nu}c_{\nu\alpha}\int\frac{d^4k}{(2\pi)^4}\frac{(k-p)_{\mu}(k-p)^{\alpha}}{[k^2-m^2](k+p)^2})\phi^*(p)\phi(-p);\nonumber\\
I_{(h)}&=&-\eta^{\mu\nu}c_{\mu\alpha}\int\frac{d^4k}{(2\pi)^4}\frac{(k-p)_{\nu}(k-p)^{\alpha}}{[k^2-m^2](k+p)^2}\phi^*(p)\phi(-p);\nonumber\\
I_{(i)}&=&\eta^{\mu\nu}c_{\alpha\beta}\int\frac{d^4k}{(2\pi)^4}\frac{(k-p)_{\mu}(k-p)_{\nu}k^{\alpha}k^{\beta}}{[k^2-m^2]^2(k+p)^2}\phi^*(p)\phi(-p).
\eea


A comment is important here. Until now, all calculations could be performed as if the gauge field propagator had not been modified. It is not the case when we consider the contribution $I_{(j)}$. As we will see, the quantum corrections on both scalar and gauge sectors will be mixed also in the divergent parts. This turns mandatory the presence of the LV modification in the gauge sector.

Let us see how the gauge propagator is modified. Initially, for the gauge field we have
\bea
\label{lag2}
{\cal L}_{gauge}=\frac{1}{4}F_{\mu\nu}F^{\mu\nu}+\frac{1}{4}\kappa_{\mu\nu\lambda\rho}F^{\mu\nu}F^{\lambda\rho}=\frac{1}{2}A_{\mu}(\eta^{\mu\nu}\Box-\partial^{\mu}\partial^{\nu}+2\kappa^{\mu\alpha\beta\nu}\partial_{\alpha}\partial_{\beta})A_{\nu},
\eea
to which we add the usual gauge-fixing term $\frac{1}{2}(\partial\cdot A)^2$, so that the gauge-field propagator, at first order in $\kappa$, becomes
$<A_{\nu}(x)A_{\rho}(y)>=-i\Box^{-1}(\eta_{\nu\rho}-\Box^{-1}2\kappa_{\nu\alpha\beta\rho}\partial^{\alpha}\partial^{\beta})\delta(x-y)$. The contribution from the diagram $(j)$ can then be written as
\bea
I_{(j)}&=&\kappa^{\alpha\beta\gamma\delta}\int\frac{d^4k}{(2\pi)^4}\frac{1}{k^4[(k-p)^2-m^2]}(2k_{\alpha}-p_{\alpha})(2k_{\delta}-p_{\delta})k_{\beta}k_{\gamma}\phi(-p)\phi^*(p).
\eea
Now we will calculate these amplitudes.

\section{Results for two-point functions}

Now, let us calculate explicitly the contributions listed above. In order to calculate the contributions $I_{a}$ to $I_{l}$, we use the Implicit Regularization framework (see a brief revision in the appendix) to isolate the divergent parts of amplitudes \cite{IR}. After a lengthy calculation, we found the following result for the vacuum polarization tensor, generated by the sum of the contributions from the diagrams $(a)\ldots (f)$,
\begin{eqnarray}
	\Pi^{\mu\nu}(p^2,m^2)&=&\frac{1}{3}\left\{p^2c^{\mu\nu}-p_{\alpha}(c^{\alpha\mu}p^{\nu}+c^{\alpha\nu}p^{\mu})+\eta^{\mu\nu}c^{\alpha\beta}p_{\alpha}p_{\beta}
	\right\}\times\nonumber\\ &\times&
	\left\{I_{log}(m^2) - \frac{b}{p^2}\left[(p^2-4 m^2)Z_0(p^2,m^2)-\frac{2}{3}p^2\right]\right\} + \nonumber\\
	&+&\frac{b}{3} c^{\alpha\beta}\frac{p_{\alpha}p_{\beta}}{p^4} (p^{\mu}p^{\nu} - p^2\eta^{\mu\nu})[6m^2 Z_0(p^2,m^2) + p^2],
\end{eqnarray}
in which $b=\frac{i}{(4\pi)^2}$. 
We note that the tensor multiplier in the first line of this equation is just the same obtained in \cite{Kost2001a} (for a symmetric $c_{\mu\nu}$) within the description of the one-loop divergent contribution to the effective action in the gauge sector of the spinor QED (that is, an essentially distinct theory). However, this is not a simple coincidence, since this is the only possible form for the self-energy tensor in first order in $c^{\mu\nu}$ consistent with the transversality requirement.
In the equation above, we used the notation of Implicit Regularization for divergent integrals,
\begin{equation}
	I_{log}(m^2)=\int^{\Lambda} \frac{d^{4}k}{(2\pi)^{4}}\frac{1}{(k^{2}-m^{2})^{2}}
\end{equation}
and
\begin{equation}
	I_{quad}(m^2)=\int^{\Lambda} \frac{d^{4}k}{(2\pi)^{4}}\frac{1}{(k^{2}-m^{2})},
\end{equation}
for logarithimic and quadratic divergences, respectively. The $\Lambda$ index indicates that the integrals are regularized and the function $Z_{0}(p^2,m^2)$ that appears in the result is a particular case of
\begin{equation}
	Z_{k}(p^2,m_1^2,m_2^2,\lambda^2)=\int_{0}^{1}dx \, x^{k}\ln\Bigg\{\frac{p^{2}x(1-x)+(m_1^2-m_2^2)x-m_1^2}{(-\lambda^{2})}\Bigg\},
	\label{z-k}
\end{equation}
in which $m_1^2=m_2^2=\lambda^2=m^2$. Straightforward checking allows to show that, for a symmetric tensor $c^{\mu\nu}$, the result is explicitly transverse. For this, an important fact is the exact cancellation of the quadratic divergences from graphs $a,b,c,d$ by the contribution from tadpole diagrams ($e$ and $f$), whose result is
\begin{equation}
	\Pi^{\mu\nu}(p^2,m^2)_{(e-f)}=I_{quad}(m^2)c^{\mu\nu}.
\end{equation}
Another important feature which comes from Implicit Regularization and that assures the cancellation of quadratic divergences and transversality is the  elimination of surface terms adopting $\alpha_i=0$ in the relations
\be
\int_k^\Lambda \frac{k_\mu k_\nu}{(k^2-m^2)^2}= \frac{\eta_{\mu\nu}}{2}\left[I_{quad}(m^2) + \alpha_1\right],
\label{TS-quad}
\ee
\be
\int_k^\Lambda \frac{k_\mu k_\nu}{(k^2-m^2)^3}= \frac{\eta_{\mu\nu}}{4}\left[I_{log}(m^2) + \alpha_2\right]
\ee
and
\be
\int_k^\Lambda \frac{k_\mu k_\nu k_\alpha k_\beta}{(k^2-m^2)^4}= \frac{\eta_{\mu\nu\alpha\beta}}{24}\left[I_{log}(m^2) + \alpha_3\right].
\ee

We can now obtain the corresponding contribution to the effective action, $\Gamma_2[A]=\frac{1}{2}A_{\mu}\Pi^{\mu\nu}A_{\nu}$, which, after transforming our result to the coordinate space, looks like 
\begin{eqnarray}
	\Gamma_2[A]&=&\frac{1}{6}c^{\alpha\beta}F_{\alpha\mu}
	\left\{I_{log}(m^2) - b\left[\left(1+4\frac{m^2}{\Box}\right)Z_0(\Box,m^2)-\frac{2}{3}\right]\right\}
	F_{\beta}^{\phantom{\beta}\mu} + \nonumber\\
	&-&\frac{b}{4}F_{\mu\nu}\frac 13c^{\alpha\beta}\frac{\partial_{\alpha}\partial_{\beta}}{\Box} \left[1-6\frac{\,\,m^2}{\Box}Z_0(\Box,m^2)\right]F^{\mu\nu},
\end{eqnarray}
where $Z_0(\Box,m^2)=\int_0^1 dx \ln\frac{m^2+\Box x(1-x)}{m^2}$.
We see that the ultraviolet divergence is concentrated in $I_{log}(m^2)$ and can be eliminated by a local counterterm $Z_cc^{\mu\nu}F_{\mu\alpha}F_{\nu}^{\phantom{\nu}\alpha}$, i.e. through a simple renormalization of the CPT-even Lorentz violating part of the photon sector, which, at the same time, requires a specific relation between $\kappa_{\mu\nu\lambda\rho}$ and $c_{\mu\nu}$, like
\begin{equation}
	\label{relconst}
	\kappa_{\mu\nu\lambda\rho}=Q[c_{\mu\lambda}\eta_{\nu\rho}-c_{\mu\rho}\eta_{\nu\lambda}+
	\eta_{\mu\lambda}c_{\nu\rho}-\eta_{\mu\rho}c_{\nu\lambda}],
\end{equation}
with $Q$ a dimensionless constant, so that one has $\frac{1}{4}\kappa_{\mu\nu\lambda\rho}F^{\mu\nu}F^{\lambda\rho}=Qc_{\mu\nu}F^{\mu\alpha}F^{\nu}_{\phantom{\nu}\alpha}$. Actually, this relation between 
	$\kappa_{\mu\nu\lambda\rho}$ and $c_{\mu\nu}$ is necessary only in order to decrease the number of independent LV parameters and simplify the form of renormalization constants, while in \cite{Kost2001a}, the renormalization constants depend on both $\kappa_{\mu\nu\lambda\rho}$ and $c_{\mu\nu}$  treated as independent parameters, and possess nontrivial tensor structures.
Aside the divergent correction, we also have UV finite nonlocal contributions, characterized by $Z_0(\Box,m^2)$, whose expansion in series in $\frac{\Box}{\,\,m^2}$, that is, the derivative expansion, yields $Z_0(\Box,m^2)=\sum\limits_{n=1}^{\infty}a_n\left(\frac{\Box}{\,\,m^2}\right)^n$, being $a_n$ dimensionless coefficients, so that we have arbitrary orders in $\Box$. We also have a finite renormalization of the Maxwell term described by the contribution $\frac{2}{3}b$, and the terms $m^2F_{\mu\nu}\Box^{-1}F^{\mu\nu}$ and $m^2F_{\mu\nu}\frac{c^{\alpha\beta}\partial_{\alpha}\partial_{\beta}}{\Box^2}F^{\mu\nu}$, which can be treated as specific gauge invariant mass terms (cf. \cite{DvaJaPi}). Actually, the first of these terms is just that one introduced in \cite{DvaJaPi}, and the second one is its further generalization involving the LV parameter.

Now, let us turn our attention to the matter sector. Following the same idea as before, we calculate the amplitudes $I_{(g)}$ to $I_{(i)}$, arriving in the following result
\begin{equation}
	\Pi(p^2,m^2)_{(g-i)}=2p_{\alpha}p_{\beta}c^{\alpha\beta}\Big\{-2I_{log}(m^2)+b\Big[\tilde{Z_{0}}+3\tilde{Z_{2}}+2p^{2}(\tilde{Y_{3}}-\tilde{Y_{4}})\Big]\Big\}
\end{equation}
in which $\tilde{Z}_k=\tilde{Z}_k(p^2,m^2) \equiv Z_k(p^2,m^2,0,m^2)$, using the definition of equation (\ref{z-k}), and $\tilde{Y}_k=\tilde{Y}_k(p^2,m^2) \equiv Y_k(p^2,m^2,0)$, given that
\begin{equation}
	Y_{k}(p^2,m_1^2,m_2^2)=\int_{0}^{1}dx\,x^{k}\frac{1}{[p^{2}x(1-x)+(m_1^2-m_2^2)x-m_1^2]}.
\end{equation}

It is important to comment on the quadratic divergences which, in principle, would appear in the individual diagrams of the matter sector. In fact, these contributions are identically null, due to the symmetry properties of the tensors $c^{\mu\nu}$ and $\kappa^{\mu\nu\alpha\beta}$. For $c^{\mu\nu}$, the would be contributions of quadratic divergences are proportional to its trace, $c^\mu_{\,\,\mu}$, which is null. Considering $\kappa^{\mu\nu\alpha\beta}$, it presents the symmetry properties of the Riemann tensor,
\be
\kappa^{\mu\nu\alpha\beta}=-\kappa^{\nu\mu\alpha\beta}=-\kappa^{\mu\nu\beta\alpha}=\kappa^{\alpha\beta\mu\nu},
\ee
and a vanishing double trace, $\kappa^{\mu\nu}_{\,\,\,\,\,\mu\nu}=0$. Since the possible contribution of quadratic divergences in this sector presents a contraction of a totally symmetric tensor with $\kappa$, it is identically null. The same argument applies to the tadpole diagrams, which do not contribute.

If we want to consider the first order in $\kappa$ only, we have the nontrivial contribution from the diagram $I_{(j)}$. So, following the same steps as in the calculations above, we obtain, for the $I_{(j)}$ integral,

\begin{equation}
	\Pi(p^2,m^2)_{(j)}= -\frac{1}{4}p_{\alpha}p_{\sigma}\kappa^{\alpha\,\,\,\beta\sigma}_{\,\,\beta}\Big\{I_{log}(m^2)-2b\tilde{Z_{1}}\Big\},
\end{equation}

In some situations, it is interesting to write the amplitude $I_{(j)}$ in terms of a four-vector $u^\mu$, using the following substitution, which is analogous to relation (\ref{relconst}):
\begin{equation}
	\kappa^{\mu\nu\alpha\beta}= Q(u^{\beta}u^{\nu}\eta^{\alpha\mu}-u^{\beta}u^{\mu}\eta^{\alpha\nu}-u^{\alpha}u^{\nu}\eta^{\beta\mu}+u^{\alpha}u^{\mu}\eta^{\beta\nu}),
\end{equation}
where $Q$ being a dimensionless constant. So, considering this form for the $\kappa^{\mu\nu\alpha\beta}$ tensor, we have the following result for amplitude $I_{(j)}$:
\begin{equation}
	\Pi(p^2,m^2)_{(j)}= -\frac{1}{4}\Big(I_{log}(m^{2})-2b\tilde{Z_{1}}\Big)\Big\{2(p \cdot u)^{2}+p^{2}u^{2}\Big\}Q.
\end{equation}
As we expected, both results are equivalent.

The complete result for the scalar field self-energy, given by the sum of the contributions on $c^{\mu\nu}$ and $\kappa^{\mu\nu\alpha\beta}$, looks like
\begin{eqnarray}
	\label{complete}
	\Pi(p^2,m^2)&=& \Pi(p^2,m^2)_{(g-i)} + \Pi(p^2,m^2)_{(j)} \nonumber \\
	&=& 2p_{\alpha}p_{\beta}c^{\alpha\beta}\Big\{-2I_{log}(m^2)+b\Big[\tilde{Z_{0}}+3\tilde{Z_{2}}+2p^{2}(\tilde{Y_{3}}-\tilde{Y_{4}})\Big]\Big\} +
	\nonumber\\ &-&
	\frac{1}{4}p_{\alpha}p_{\beta}\kappa^{\alpha\,\,\,\gamma\beta}_{\,\,\,\gamma}\Big\{I_{log}(m^{2})-2b\tilde{Z_{1}}\Big\}.
\end{eqnarray}
The corresponding contribution to the effective action can be written as $\Gamma[\phi]=\frac{1}{2}\int\frac{d^4p}{(2\pi)^4}\phi^*(-p)\Pi(p)\phi(p)$. Effectively we can write $\Pi(p)\equiv \tilde{\Pi}^{\alpha\beta}(p)p_{\alpha}p_{\beta}$, in which $\tilde{\Pi}^{\alpha\beta}(p)$ can be read off from (\ref{complete}):
\be\label{Eqc}
\Pi^{\alpha\beta}=2c^{\alpha\beta}\Big\{-2I_{log}(m^2)+b\Big[\tilde{Z_{0}}+3\tilde{Z_{2}}+2p^{2}(\tilde{Y_{3}}-\tilde{Y_{4}})\Big]\Big\}-
\frac{1}{4}\kappa^{\alpha\gamma\beta}_{\gamma}\Big\{I_{log}(m^{2})-2b\tilde{Z_{1}}\Big\}.
\ee
We note that, substituting (\ref{relconst}) into (\ref{Eqc}), we can present it as follows:
\be\label{Eqc1}
\Pi^{\alpha\beta}=2c^{\alpha\beta}\Big\{-I_{log}(m^2)\Big(2+Q/2\Big)+b\Big[\tilde{Z_{0}}+Q\tilde{Z_{1}}+3\tilde{Z_{2}}+2p^{2}(\tilde{Y_{3}}-\tilde{Y_{4}})\Big]\Big\}.
\ee
The $\Pi^{\alpha\beta}$ can be expanded in power series in the momentum $p^2$. So, our effective action, being rewritten in the coordinate space, can be presented in the form
\begin{equation}
	\Gamma[\phi]=\int d^4 x \partial_{\alpha}\phi^*\Pi^{\alpha\beta}\partial_{\beta}\phi,
\end{equation}
where we can write $\Pi^{\alpha\beta}=c^{\alpha\beta}\hat{L}_1$, with $\hat{L}_1$ being a single scalar operator which can be represented as power series in derivatives. The lower order term of this expansion will yield the aether-like structure $\int d^4 x \partial_{\alpha}\phi^*c^{\alpha\beta}\partial_{\beta}\phi$, in which $\Pi^{\alpha\beta}$ is completely described in terms of $c^{\alpha\beta}$ parameter. The higher order terms of the expansion will correspond to various powers of $\frac{p^2}{m^2}$, arising from expansions of various $\tilde{Z}_i$ and $\tilde{Y}_i$.

\section{Conclusions}

Let us discuss our results. Within this paper, we started with the simplest CPT-even Lorentz-violating extension of scalar QED, where LV parameters are introduced both in scalar and gauge sectors. In this theory, we calculated the one-loop contributions to the two-point function, both in gauge and scalar sectors. We found that aether-like structures arise in both cases. While our results are divergent, their renormalization can be performed through simple wave function renormalizations for aether terms in both sectors. In principle, it is natural to expect that the renormalized triple and quartic vertices will allow to write the one-loop result in the scalar sector in the form $\int d^4 x (D_{\alpha}\phi)^*\tilde{c}^{\alpha\beta}D_{\beta}\phi$, in which $\bar{c}^{\alpha\beta}$ is some symmetric tensor. However, to confirm this, one needs results for three-and four-point scalar-vector functions, in order to study the renormalization of the coupling. This calculation will be performed in a separate paper.

It is natural to expect that our results can be used for other studies, for example, for detailed discussions of the Higgs mechanism, generalizing the results of \cite{Scarp2013} to the CPT-even case. We plan to do it in a forthcoming paper.

\vspace{0.5cm}

{\bf Acknowledgments.}  Authors are grateful to J. R. Nascimento for important discussions. A.P.B.S acknowledges CNPq by financial support. The work by A. Yu. P. has been partially supported by the
CNPq project No. 301562/2019-9.

\section{Appendix}

We present below a brief review of Implicit Regularization based in the paper \cite{Zanotelli}. The traditional procedure can be formulated by a set of rules, the first one being the assumption of a regularization acting in the complete amplitude. This assures algebraic manipulations can be carried out in the integrand. The group algebra, then, is performed and the momentum-space amplitude is written as a combination of basic integrals, like, for example,
\be
\label{Integrals}
I,I_\mu, I_{\mu\nu}= \int^\Lambda \frac{d^4k}{(2\pi)^4}\frac{1,k_\mu,k_\mu k_\nu}{(k^2-m^2)[(p-k)^2-m^2]},
\ee
multiplied by polynomials of the external momentum and typical objects of the symmetry group. The index $\Lambda$ in the integrals indicates they are regularized. Each one of these basic integrals can be treated following a set of rules and a table with their results can be used whenever a new 	calculation is being performed.

In order to obtain the divergent part of a basic integral, the identity,
\bq
\frac {1}{(p-k)^2-m^2}=\frac{1}{(k^2-m^2)} - \frac{p^2-2p \cdot k}{(k^2-m^2)
	\left[(p-k)^2-m^2\right]},
\label{ident}
\eq
is applied recursively, so as the divergent part do not have the external momentum $p$ in the denominator. The remaining divergent integrals have the form
\be
\int_k ^\Lambda \frac{k_{\mu_1}k_{\mu_2}\cdots}{(k^2-m^2)^\alpha},
\ee
in which $\int_k$ means $\int d^4k/(2 \pi)^4$. The divergent integrals with Lorentz indices must be expressed in terms of divergent scalar integrals and surface terms. For example:
\be
\int_k ^\Lambda \frac{k_\mu k_\nu}{(k^2-m^2)^3}=
\frac 14 \left\{ \eta_{\mu \nu} \int_k^\Lambda
\frac {1}{(k^2-m^2)^2} -\int_k^\Lambda \frac{\partial}{\partial k^\nu}
\left( \frac{k_\mu}{(k^2-m^2)^2}\right)\right\}.
\label{ts}
\ee
The surface terms, that vanish for finite integrals, depend here on the method of regularization in use. They are symmetry-violating terms, since the possibility of making shifts in the integrals needs the surface terms to vanish. Non-null surface terms imply that the amplitude depend on the momentum routing choice. 

Finally, the divergent part of the integrals is written as a combination of the basic divergences
\be
I_{log}(m^2)=\int_k^\Lambda \frac{1}{(k^2-m^2)^2}
\,\,\,\,\,\,\,\,  \mbox{and}  \,\,\,\,\,\,\,\
I_{quad}(m^2)=\int_k^\Lambda \frac{1}{(k^2-m^2)},
\label{basicdiv}
\ee
which will require local counterterms in the process of renormalization.

The version of Implicit Regularization (IReg) in which all the surface terms are fixed null from the beginning is known as Constrained Implicit Regularization (CIReg). Recently a new procedure for implementing this constrained version, which greatly simplifies the calculations, was developed \cite{Zanotelli}. As in the original procedure, a regularization scheme is assumed to be acting in the complete amplitude. Then Feynman parametrization is applied to the integrals before separating the divergent parts. It can be applied to the complete amplitude, such that the needed shift in the momentum of integration after Feynman parametrization is just a modification in the loop momentum. The algebra of the group of symmetry is then carried out. Note that when the integrals are treated separately, the algebra is performed before Feynman parametrization.

The integrals in the momenta are separated by degree of divergence, all with even powers of the integration momentum in the numerator, of the type
\be
\int^\Lambda\frac{d^4k}{(2\pi)^4}\frac{1, k_\mu k_\nu, k_\mu k_{\nu} k_{\alpha}k_{\beta},\cdots}{(k^2+H^2)^n},
\ee
being $H^2$ function of the external momenta, of the masses and of the Feynman parameters. Care must be taken when factors of $k^2$ appear in the numerator: they should be canceled with factors in the denominator by adding and subtracting $H^2$. As before, for the divergent parts, the surface terms are eliminated from the divergent integrals with Lorentz indices. For one-loop logarithmically and quadratically divergent integrals, it is used, respectively,
\be
\int^\Lambda\frac{d^4k}{(2\pi)^4}\frac{k_{\mu_1} k_{\mu_2} \cdots k_{\mu_n}}{(k^2+H^2)^{2+\frac n2}}=
\frac{\eta_{\mu_1 \mu_2 \cdots \mu_n}}{2^{\frac n2} (\frac n2 +1)!}
\int^\Lambda\frac{d^4k}{(2\pi)^4}\frac{1}{(k^2+H^2)^2} 
\label{RC-geral}
\ee
and
\be
\int^\Lambda\frac{d^4k}{(2\pi)^4}\frac{k_{\mu_1} k_{\mu_2} \cdots k_{\mu_n}}{(k^2+H^2)^{1+\frac n2}}=
\frac{\eta_{\mu_1 \mu_2 \cdots \mu_n}}{2^{\frac n2} (\frac n2)!} 
\int^\Lambda\frac{d^4k}{(2\pi)^4}\frac{1}{(k^2+H^2)},
\label{RC-geral-2}
\ee
in which $n$ is even and $\eta_{\mu_1 \mu_2 \cdots \mu_n}$ is the symmetric combination of the products of metric tensors, $\eta_{\mu_1 \mu_2}\cdots \eta_{\mu_{n-1}\mu_n}$, with coefficient $1$. In order to obtain the relations above, it is used recursively the relation,
\be
\int_k^\Lambda \frac{\partial}{\partial k^{\mu_n}}\left(\frac{k_{\mu_1}\cdots k_{\mu_{n-1}}}{(k^2+H^2)^{m-1}}\right) =
\int_k^\Lambda \frac{{\cal S}[\eta_{\mu_1 \mu_n}k_{\mu_2}\cdots k_{\mu_{n-1}}]}{(k^2+H^2)^{m-1}}
- 2 (m-1) \int_k^\Lambda\frac{k_{\mu_1} \cdots k_{\mu_n}}{(k^2+H^2)^m},
\ee
until the first integral of the second member of the equation is scalar. Apart two integrals, including the scalar one, all the others will be surface terms which are gathered in one parameter and fixed null. In the equation above, ${\cal S}[T_{\mu_1 \cdots \mu_n}]$ means the minimal symmetrization of the tensor $T$, as in the example ${\cal S}[k_\mu k_\nu p_\alpha]= k_\mu k_\nu p_\alpha + k_\mu k_\alpha p_\nu + k_\nu k_\alpha p_\mu$. By the definitions of equation (\ref{basicdiv}), the remaining scalar divergences above are $I_{log}(-H^2)$ and $I_{quad}(-H^2)$.

Next, algebraic identities are used in order to get the divergent integrals free from the external momenta, with the recursive use of the simpler expansion,
\be
\frac{1}{(k^2+H^2)}= \frac{1}{(k^2-\lambda^2)} - \frac{\lambda^2+H^2}{(k^2-\lambda^2)(k^2+H^2)},
\ee
such that closed expressions to be used in any calculation are obtained:
\be
I_{log}(-H^2)= I_{log}(\lambda^2) - \frac{i}{16\pi^2}\ln{\left(-\frac{H^2}{\lambda^2}\right)}
\label{scale-1}
\ee
and
\be
I_{quad}(-H^2)=I_{quad}(\lambda^2)-(\lambda^2+H^2)I_{log}(\lambda^2) - \frac{i}{16\pi^2} 
\left[\lambda^2 + H^2 - H^2 \ln{\left(-\frac{H^2}{\lambda^2}\right)} \right].
\label{scale-2}
\ee
These scale relations, as a byproduct, introduce an energy scale for the renormalization group, $\lambda^2$. The basic divergences are factorized out of the integrals in the Feynman parameters, which can be computed. The divergent part of the amplitudes is then written in terms of the basic divergences, $I_{log}(\lambda^2)$, $I_{quad}(\lambda^2)$, etc.

The implementation of the above steps simplifies a lot the calculations of the finite parts. An additional advantage is related to models which present fields with different masses or non-massive fields, since all the mass dependence is inside $H^2$.




\begin{thebibliography}{99}
	
	\bibitem{kostelecky1}  D. Colladay and V. A. Kostelecky, Phys. Rev. D55, 6760 (1997), hep-ph/9703464.
	
	\bibitem{kostelecky2}  D. Colladay and V. A. Kostelecky, Phys. Rev. D58, 116002 (1998), hep-ph/9809521.
	
	\bibitem{JK} R. Jackiw, V. A. Kostelecky, Phys. Rev. Lett. 82, 3572 (1999), hep-ph/9901358.
	
	\bibitem{ourrev} A.~F.~Ferrari, J.~R.~Nascimento and A.~Y.~Petrov,
	Eur. Phys. J. C \textbf{80} (2020) 459,
	arXiv:1812.01702 [hep-th].
	
	\bibitem{Kostelecky:1988zi} V.~A.~Kostelecky and S.~Samuel, 
	Phys. Rev. D \textbf{39} (1989), 683.
	
	\bibitem{Doplicher:1994tu} S.~Doplicher, K.~Fredenhagen and J.~E.~Roberts, 
	Commun. Math. Phys. \textbf{172} (1995), 187-220.
	
	\bibitem{Bojowald:2004bb} M.~Bojowald, H.~A.~Morales-Tecotl and H.~Sahlmann, 
	Phys. Rev. D \textbf{71} (2005), 084012.
	\bibitem{Amelino-Camelia:2008aez} G.~Amelino-Camelia, 
	Living Rev. Rel. \textbf{16} (2013), 5
	
	\bibitem{Addazi:2021xuf} 
	A.~Addazi, J.~Alvarez-Muniz, R.~A.~Batista, G.~Amelino-Camelia, V.~Antonelli, M.~Arzano, M.~Asorey, J.~L.~Atteia, S.~Bahamonde and F.~Bajardi, \textit{et al.} 
	Prog. Part. Nucl. Phys. (2022) 103948. [arXiv:2111.05659 [hep-ph]]. 
	
	\bibitem{Kostelecky:2016kfm} 
	V.~A.~Kosteleck\'y and M.~Mewes, 
	Phys. Lett. B \textbf{757} (2016), 510-514
	
	\bibitem{Kostelecky:2016kkn} V.~A.~Kosteleck\'y, A.~C.~Melissinos and M.~Mewes, 
	Phys. Lett. B \textbf{761} (2016), 1-7.
	
	\bibitem{Schreck:2016qiz} M.~Schreck, 
	Class. Quant. Grav. \textbf{34} (2017), 135009.
	
	\bibitem{Kostelecky:2015dpa} V.~A.~Kosteleck\'y and J.~D.~Tasson, 
	Phys. Lett. B \textbf{749} (2015), 551-559.
	
	\bibitem{Altschul:2008qg} B.~Altschul, 
	Phys. Rev. D \textbf{78} (2008), 085018.
	
	\bibitem{Bertolami:1999da} O.~Bertolami and C.~S.~Carvalho, 
	Phys. Rev. D \textbf{61} (2000), 103002.
	
	\bibitem{Kostelecky:2008ts} V.~A.~Kostelecky and N.~Russell, ``Data Tables for Lorentz and CPT Violation,'' [arXiv:0801.0287 [hep-ph]].
	
	\bibitem{Kost2001a} V. A. Kostelecky, C. D. Lane, A. G. M. Pickering,
	Phys. Rev. D65, 056006 (2002), hep-th/0111123.
	
	\bibitem{Scarp1} A.~P.~Baeta Scarpelli, L.~C.~T.~Brito, J.~C.~C.~Felipe, J.~R.~Nascimento and A.~Y.~Petrov,
	EPL \textbf{123} (2018), 21001,
	arXiv:1805.06256.
	
	\bibitem{Scarp2} L.~C.~T.~Brito, J.~C.~C.~Felipe, J.~R.~Nascimento, A.~Y.~Petrov and A.~P.~B.~Scarpelli,
	Phys. Rev. D \textbf{102} (2020) 075017,
	arXiv:2007.11538.
	
	\bibitem{AltHiggs} B.~Altschul,
	Phys. Rev. D \textbf{86} (2012), 045008
	[arXiv:1202.5993 [hep-th]].
	
	\bibitem{Scarp2013} L.~C.~T.~Brito, H.~G.~Fargnoli and A.~P.~Ba\^eta Scarpelli,
	Phys. Rev. D \textbf{87} (2013), 125023,
	arXiv:1304.6016.
	
	\bibitem{effpotLV} A.~P.~Baeta Scarpelli, L.~C.~T.~Brito, J.~C.~C.~Felipe, J.~R.~Nascimento and A.~Y.~Petrov,
	Eur. Phys. J. C \textbf{77} (2017), 850,
	arXiv:1704.08556.
	
	\bibitem{Hel} A. P. Baeta Scarpelli, H. Belich, J. L. Boldo, J. A. Helayel-Neto, Phys. Rev. D 67,  085021 (2003).
	
	\bibitem{IR} O. A. Battistel, A. L. Mota, M. C. Nemes, Mod. Phys. Lett. A \textbf{13}, 1597 (1998);
	O. A. Battistel, M. C. Nemes, Phys. Rev. D \textbf{59}, 055010 (1999), hep-th/9811154;
	A. L Cherchiglia, M. Sampaio, M. C. Nemes, Int. J. Mod. Phys. A \textbf{26}, 2591 (2011), arXiv: 1008.1377.
	
	\bibitem{Petrov} A. P. Baeta Scarpelli, T. Mariz, J. R. Nascimento, A. Yu. Petrov, Eur. Phys. J. C \textbf{73}, 2526 (2013), arXiv: 1304.2256.
	
	\bibitem{DvaJaPi} G.~Dvali, R.~Jackiw and S.~Y.~Pi,
	Phys. Rev. Lett. \textbf{96} (2006), 081602,
	arXiv: hep-th/0511175.
	
	\bibitem{Zanotelli} B. Z. Felippe, A. P. Baêta Scarpelli, A. R. Vieira, J. C. C. Felipe, \textit{Advances in 4D regularizations towards the systematization of calculations in multiple loops}, arXiv: 2112.03844.
	
\end{thebibliography}
\end{document}